\documentclass[12pt]{article}

\usepackage{graphicx}    % standard LaTeX graphics tool
\usepackage{url}

\def\beq{\begin{equation}}
\def\eeq{\end{equation}}
\def\nn{\nonumber}
\def\bea{\begin{eqnarray}}
\def\eea{\end{eqnarray}}
\def\ba{\begin{array}}                  %array
\def\ea{\end{array}}

\begin{document}

\begin{center}
\noindent{\LARGE \bf Finite Unified Theories and the Higgs Mass
  Prediction}
\footnote{Based on invited lectures
    given at a) II Summer School in Modern Mathematical Physics,
    Kapaonik 2002, b) 2nd ISPM Workshop on Particles and Cosmology,
    Modern Trends in Gravity, Cosmology and Particle Physics, Tbilisi
    2002, c) Yerevan Stepanakert 2003 Workshop, Yerevan 2003, d)
    Recent Developments in String/M-theory and Field Theory (36th
    Ahrenshoop Symposium), Schmockitz-Berlin area 2003, e) 9th
    Adriatic Meeting, Particle Physics and the Universe, Dubrovnik
    2003, f) Matter to the Deepest: Recent Developments in Physics of
    Fundamental Interactions, Ustron 2003, g) 2nd Aegean Summer School
    on the Early Universe, Syros 2003.}
%\end{center}
\\[15pt]
{\large Abdelhak Djouadi$^a$, Sven Heinemeyer$^b$,  Myriam
  Mondrag{\'o}n$^c$ \\[3pt] 
and George Zoupanos$^d$}\\[15pt]
{\small 
 $^a$Laboratoire de Physique Math{\'e}matique et Th{\'e}orique,\\[3pt]
  Universit{\'e} de Montpellier II, France\\[3pt]
  \texttt{abdelhak.djouadi@cern.ch}  \\[7pt]
$^b$ Dept. of Physics, CERN, TH
  Division, 1211 Geneva 23, Switzerland \\[3pt]
  \texttt{sven.heinemeyr@cern.ch}\\[7pt]
 $^c$ Instituto de F{\'\i}sica, UNAM,
  Apdo. Postal 20-364, M{\'e}xico 01000, D.F., M{\'e}xico\\[3pt]
  \texttt{myriam@fisica.unam.mx}\\[7pt]
$^d$ Physics Dept., Nat. Technical
  University, GR-157 80 Zografou, Athens, Greece\\[3pt]
  \texttt{george.zoupanos@cern.ch}
}
\end{center}

\begin{abstract}  
  Finite Unified Theories (FUTs) are N=1 supersymmetric Grand Unified
  Theories, which can be made all-loop finite, both in the
  dimensionless (gauge and Yukawa couplings) and dimensionful (soft
  supersymmetry breaking terms) sectors.  This remarkable property
  provides a drastic reduction in the number of free parameters, which
  in turn leads to an accurate prediction of the top quark mass in the
  dimensionless sector, and predictions for the Higgs boson mass and
  the supersymmetric spectrum in the dimensionful sector.  Here we
  examine the predictions of two FUTs taking into account a number of
  theoretical and experimental constraints. For the first one we
  present the results of a detailed scanning concerning the Higgs mass
  prediction, while for the second we present a representative
  prediction of its spectrum.
\end{abstract}

\section{Introduction}
Finite Unified Theories are $N=1$ supersymmetric Grand Unified
Theories (GUTs) which can be made finite even to all-loop orders,
including the soft supersymmetry breaking sector.  The method to
construct GUTs with reduced independent parameters
\cite{zoup-kmz1,zoup-zim1} consists of searching for renormalization
group invariant (RGI) relations holding below the Planck scale, which
in turn are preserved down to the GUT scale. Of particular interest is
the possibility to find RGI relations among couplings that guarantee
finitenes to all-orders in perturbation theory
\cite{zoup-lucchesi1,zoup-ermushev1}.  In order to achieve the latter
it is enough to study the uniqueness of the solutions to the one-loop
finiteness conditions \cite{zoup-lucchesi1,zoup-ermushev1}.  The
constructed {\it finite unified} $N=1$ supersymmetric SU(5) GUTs, using
the above tools, predicted correctly from the dimensionless sector
(Gauge-Yukawa unification), among others, the top quark mass
\cite{zoup-finite1}.  The search for RGI relations and finiteness has
been extended to the soft supersymmetry breaking sector (SSB) of these
theories \cite{zoup-kmz2,zoup-jack2}, which involves parameters of
dimension one and two.  Eventually, the full theories can be made
all-loop finite and their predictive power is extended to the Higgs
sector and the supersymmetric spectrum (s-spectrum).  The purpose of
the present article is to start an exhaustive search of these latter
predictions, as well as to provide a rather dense review of the
subject.

\section{Reduction of Couplings and Finiteness in $N=1$ SUSY Gauge Theories}

Here let us review the main points and ideas concerning the {\it reduction
of couplings} and {\it finiteness} in $N=1$ supersymmetric theories. 
A RGI relation among couplings $g_i$, $ \Phi (g_1,\cdots,g_N) ~=~0, $ has to
satisfy the partial differential equation $ \mu~ d \Phi /d \mu ~=~
\sum_{i=1}^{N} \,\beta_{i}\,\partial \Phi /\partial g_{i}~=~0, $ where $\beta_i$ is the
$\beta$-function of $g_i$.  There exist ($N-1$) independent $\Phi$'s, and
finding the complete set of these solutions is equivalent to solve the
so-called reduction equations (REs) \cite{zoup-zim1}, $ \beta_{g} \,(d
g_{i}/d g) =\beta_{i}~,~i=1,\cdots,N, $ where $g$ and $\beta_{g}$ are the
primary coupling and its $\beta$-function.  Using all the $(N-1)\,\Phi$'s
to impose RGI relations, one can in principle express all the
couplings in terms of a single coupling $g$.  The complete reduction,
which formally preserves perturbative renormalizability, can be
achieved by demanding a power series solution, whose uniqueness can be
investigated at the one-loop level.
 
Finiteness can be understood by considering a chiral, anomaly free,
~~$N=1$ globally supersymmetric gauge theory based on a group G with
gauge coupling constant $g$. The superpotential of the theory is given
by
\begin{equation}
 W= \frac{1}{2}\,m^{ij} \,\Phi_{i}\,\Phi_{j}+
\frac{1}{6}\,C^{ijk} \,\Phi_{i}\,\Phi_{j}\,\Phi_{k}~, 
\label{supot}
\end{equation}
where $m^{ij}$ (the mass terms) and $C^{ijk}$ (the Yukawa couplings)
are gauge invariant tensors and the matter field $\Phi_{i}$ transforms
according to the irreducible representation $R_{i}$ of the gauge group
$G$.  

 The one-loop $\beta$-function of the gauge
coupling $g$ is given by 
\bea
\beta^{(1)}_{g}&=&\frac{d g}{d t} =
\frac{g^3}{16\pi^2}\,[\,\sum_{i}\,l(R_{i})-3\,C_{2}(G)\,]~,
\label{betag}
\eea
where $l(R_{i})$ is the Dynkin index of $R_{i}$ and $C_{2}(G)$
 is the
quadratic Casimir of the adjoint representation of the
gauge group $G$. The $\beta$-functions of
$C^{ijk}$,
by virtue of the non-renormalization theorem, are related to the
anomalous dimension matrix $\gamma^j_i$ of the matter fields
$\Phi_{i}$ as:
\beq
\beta_{C}^{ijk}=\frac{d}{dt}\,C^{ijk}
=C^{ijp}\,
\sum_{n=1}\frac{1}{(16\pi^2)^n}\,\gamma_{p}^{k(n)} +(k
\leftrightarrow i) +(k\leftrightarrow j)~.
\label{betay}
\eeq
At one-loop level $\gamma^j_i$ is given by 
\bea
\gamma_i^{j(1)}=\frac{1}{2}C_{ipq}\,C^{jpq}-2\,g^2\,C_{2}(R_{i})\delta_i^j~,
\label{gamay}
\eea
where $C_{2}(R_{i})$ is the quadratic Casimir of the representation
$R_{i}$, and $C^{ijk}=C_{ijk}^{*}$.

All the one-loop $\beta$-functions of the theory vanish if the
$\beta$-function of the gauge coupling $\beta_g^{(1)}$, and the
anomalous dimensions $\gamma_i^{j(1)}$,
vanish, i.e.
\begin{equation}
\sum _i \ell (R_i) = 3 C_2(G) \,,~
\frac{1}{2}C_{ipq} C^{jpq} = 2\delta _i^j g^2  C_2(R_i)\ ,
\label{zoup-fini}
\end{equation}
where $l(R_i)$ is the Dynkin index of $R_i$, and $C_2(G)$ is the
quadratic Casimir invariant of the adjoint representation of $G$.

A very interesting result is that the conditions (\ref{zoup-fini}) are
necessary and sufficient for finiteness at
the two-loop level \cite{soft,zoup-jack1}.

The one- and two-loop finiteness conditions (\ref{zoup-fini}) restrict
considerably the possible choices of the irreducible
representations $R_i$ for a given group $G$ as well as the Yukawa
couplings in the superpotential (\ref{supot}).  Note in particular
that the finiteness conditions cannot be applied to the supersymmetric
standard model (SSM), since the presence of a $U(1)$ gauge group is
incompatible with the condition (\ref{zoup-fini}), due to
$C_2[U(1)]=0$.  This leads to the expectation that finiteness should
be attained at the grand unified level only, the SSM being just the
corresponding low-energy, effective theory.

The finiteness conditions impose relations between gauge and Yukawa
couplings.  Therefore, we have to guarantee that such relations
leading to a reduction of the couplings hold at any renormalization
point.  The necessary, but also sufficient, condition for this to
happen is to require that such relations are solutions to the
reduction equations (REs) to all orders.  The all-loop order
finiteness theorem of ref.\cite{zoup-lucchesi1} is based on: (a) the
structure of the supercurrent in $N=1$ SYM and on (b) the
non-renormalization properties of $N=1$ chiral anomalies
\cite{zoup-lucchesi1}.  Alternatively, similar results can be obtained
\cite{zoup-ermushev1,zoup-strassler} using an analysis of the all-loop
NSVZ gauge beta-function \cite{zoup-novikov1}.

\section{Soft supersymmetry breaking and finiteness}

The above described method of reducing the dimensionless couplings has
been extended \cite{zoup-kmz2,zoup-jack2} to the soft supersymmetry
breaking (SSB) dimensionful parameters of $N=1$ supersymmetric
theories.  Recently  very interesting progress has been made
\cite{zoup-avdeev1}-\cite{zoup-kkmz1} concerning the renormalization
properties of the SSB parameters, based conceptually and technically on
the work of ref.~\cite{zoup-yamada1}.  In this work the powerful
supergraph method \cite{zoup-delbourgo1} for studying supersymmetric
theories has been applied to the softly broken ones by using the
``spurion'' external space-time independent superfields
\cite{zoup-girardello1}.  In the latter method a softly broken
supersymmetric gauge theory is considered as a supersymmetric one in
which the various parameters such as couplings and masses have been
promoted to external superfields that acquire ``vacuum expectation
values''. Based on this method the relations among the soft term
renormalization and that of an unbroken supersymmetric theory have
been derived. In particular the $\beta$-functions of the parameters of
the softly broken theory are expressed in terms of partial
differential operators involving the dimensionless parameters of the
unbroken theory. The key point in the strategy of
refs.~\cite{zoup-avdeev1}-\cite{zoup-kkmz1} in solving the set of
coupled differential equations so as to be able to express all
parameters in a RGI way, was to transform the partial differential
operators involved to total derivative operators \cite{zoup-avdeev1}.
It is indeed possible to do this on the RGI surface which is defined
by the solution of the reduction equations.  In addition it was found
that RGI SSB scalar masses in Gauge-Yukawa unified models satisfy a
universal sum rule at one-loop \cite{zoup-kkk1}. This result was
generalized to two-loops for finite theories \cite{zoup-kkmz1}, and
then to all-loops for general Gauge-Yukawa and Finite Unified Theories
\cite{zoup-kkz}. 

In order to obtain a feeling of some of the above
results, consider the superpotential given by (1) along with the
Lagrangian for SSB terms 
\begin{equation}
%\begin{split}  
-{\cal L}_{\rm SB} = \frac{1}{6}
\,h^{ijk}\,\phi_i \phi_j \phi_k +
\frac{1}{2} \,b^{ij}\,\phi_i \phi_j \\
+ \frac{1}{2} \,(m^2)^{j}_{i}\,\phi^{*\,i} \phi_j+ \frac{1}{2} \,M\,\lambda
\lambda+\mbox{H.c.},
%\end{split} 
\end{equation}
where the $\phi_i$ are the scalar parts of the
chiral superfields $\Phi_i$ , $\lambda$ are the gauginos and $M$ their
unified mass.  Since only finite theories are considered here,
it is assumed that the gauge group is a simple group and the
one-loop $\beta$-function of the gauge coupling $g$ vanishes. It is also
assumed that the reduction equations admit power series solutions of
the form 
\beq
C^{ijk} = g\,\sum_{n=0}\,\rho^{ijk}_{(n)} g^{2n}~.  
\eeq
According
to the finiteness theorem \cite{zoup-lucchesi1}, the theory is then finite
to all-orders in perturbation theory, if, among others, the one-loop
anomalous dimensions $\gamma_{i}^{j(1)}$ vanish.  The one- and two-loop
finiteness for $h^{ijk}$ can be achieved by \cite{zoup-jack1}
\beq
 h^{ijk} = -M C^{ijk}+\dots =-M
\rho^{ijk}_{(0)}\,g+O(g^5)~.
\eeq

An additional constraint in the SSB sector up to two-loops
\cite{zoup-kkmz1}, concerns the soft scalar masses as follows
\begin{equation}
\frac{(~m_{i}^{2}+m_{j}^{2}+m_{k}^{2}~)}{M M^{\dag}} =
1+\frac{g^2}{16 \pi^2}\,\Delta^{(2)}
+O(g^4)~
\label{zoup-sumr}
\end{equation}
for i, j, k with $\rho^{ijk}_{(0)} \neq 0$, where $\Delta^{(2)}$ is
the two-loop correction
\begin{equation}
\Delta^{(2)} =  -2\sum_{l} [(m^{2}_{l}/M M^{\dag})-(1/3)]~T(R_l),
\end{equation}
which vanishes for the
universal choice \cite{zoup-jack1}, i.e. when all the soft scalar
masses are the same at the unification point.

If we know higher-loop $\beta$-functions explicitly, we can follow the same 
procedure and find higher-loop RGI relations among SSB terms.
However, the $\beta$-functions of the soft scalar masses are explicitly
known only up to two loops.
In order to obtain higher-loop results, we need something else instead of 
knowledge of explicit $\beta$-functions, e.g. some relations among 
$\beta$-functions.

The recent progress made using the spurion technique
\cite{zoup-delbourgo1,zoup-girardello1} leads to
the following  all-loop relations among SSB $\beta$-functions, 
\cite{zoup-avdeev1}-\cite{zoup-kkmz1}
\bea
\beta_M &=& 2{\cal O}\left({\beta_g\over g}\right)~,
\label{betaM}\\
\beta_h^{ijk}&=&\gamma^i{}_lh^{ljk}+\gamma^j{}_lh^{ilk}
+\gamma^k{}_lh^{ijl}\nn\\
&&-2\gamma_1^i{}_lC^{ljk}
-2\gamma_1^j{}_lC^{ilk}-2\gamma_1^k{}_lC^{ijl}~,\\
(\beta_{m^2})^i{}_j &=&\left[ \Delta 
+ X \frac{\partial}{\partial g}\right]\gamma^i{}_j~,
\label{betam2}\\
{\cal O} &=&\left(Mg^2{\partial\over{\partial g^2}}
-h^{lmn}{\partial
\over{\partial C^{lmn}}}\right)~,
\label{diffo}\\
\Delta &=& 2{\cal O}{\cal O}^* +2|M|^2 g^2{\partial
\over{\partial g^2}} +\tilde{C}_{lmn}
{\partial\over{\partial
C_{lmn}}} +\tilde{C}^{lmn}{\partial\over{\partial C^{lmn}}}~,
\eea
where $(\gamma_1)^i{}_j={\cal O}\gamma^i{}_j$, 
$C_{lmn} = (C^{lmn})^*$, and 
\bea
\tilde{C}^{ijk}&=&
(m^2)^i{}_lC^{ljk}+(m^2)^j{}_lC^{ilk}+(m^2)^k{}_lC^{ijl}~.
\label{tildeC}
\eea
It was also found \cite{zoup-jack4}  that the relation
\bea
h^{ijk} &=& -M (C^{ijk})'
\equiv -M \frac{d C^{ijk}(g)}{d \ln g}~,
\label{h2}
\eea
among couplings is all-loop RGI. Furthermore, using the all-loop gauge
$\beta$-function of Novikov {\em et al.} 
\cite{zoup-novikov1} given
by 
\bea
\beta_g^{\rm NSVZ} &=& 
\frac{g^3}{16\pi^2} 
\left[ \frac{\sum_l T(R_l)(1-\gamma_l /2)
-3 C(G)}{ 1-g^2C(G)/8\pi^2}\right]~, 
\label{bnsvz}
\eea 
it was found the all-loop RGI sum rule \cite{zoup-kkz},
\bea
m^2_i+m^2_j+m^2_k &=&
|M|^2 \{~
\frac{1}{1-g^2 C(G)/(8\pi^2)}\frac{d \ln C^{ijk}}{d \ln g}
+\frac{1}{2}\frac{d^2 \ln C^{ijk}}{d (\ln g)^2}~\}\nn\\
& &+\sum_l
\frac{m^2_l T(R_l)}{C(G)-8\pi^2/g^2}
\frac{d \ln C^{ijk}}{d \ln g}~.
\label{sum2}
\eea
In addition 
the exact-$\beta$-function for $m^2$
in the NSVZ scheme has been obtained \cite{zoup-kkz} for the first time and
is given by
\bea
\beta_{m^2_i}^{\rm NSVZ} &=&\left[~
|M|^2 \{~
\frac{1}{1-g^2 C(G)/(8\pi^2)}\frac{d }{d \ln g}
+\frac{1}{2}\frac{d^2 }{d (\ln g)^2}~\}\right.\nn\\
& &\left. +\sum_l
\frac{m^2_l T(R_l)}{C(G)-8\pi^2/g^2}
\frac{d }{d \ln g}~\right]~\gamma_{i}^{\rm NSVZ}~.
\label{bm23}
\eea

\section{Finite Unified Theories}

In this section we examine two concrete $SU(5)$ finite models, where
the reduction of couplings in the dimensionless and dimensionful
sector has been achieved.  A predictive Gauge-Yukawa unified $SU(5)$
model which is finite to all orders, in addition to the requirements
mentioned already, should also have the following properties:

\begin{enumerate}

\item 
One-loop anomalous dimensions are diagonal,
i.e.,  $\gamma_{i}^{(1)\,j} \propto \delta^{j}_{i} $.
\item Three fermion generations, in the irreducible representations
  $\overline{\bf 5}_{i},{\bf 10}_i~(i=1,2,3)$, which obviously should
  not couple to the adjoint ${\bf 24}$.
\item The two Higgs doublets of the MSSM should mostly be made out of a
pair of Higgs quintet and anti-quintet, which couple to the third
generation.
\end{enumerate}

In the following we discuss two versions of the all-order finite
model.  The model of ref.~\cite{zoup-finite1}, which will be labeled
${\bf A}$, and a slight variation of this model (labeled ${\bf B}$) ,
which can also be obtained from the class of the models suggested by
Kazakov {\em et al.} \cite{zoup-avdeev1} with a modification to
suppress non-diagonal anomalous dimensions\footnote{An extension to three
  families, and the generation of quark mixing angles and masses in
  Finite Unified Theories has been addressed in \cite{zoup-Babu:2002in},
  where several realistic examples are given. These extensions are not 
considered here.}.

The  superpotential which describes the two models 
takes the form \cite{zoup-finite1,zoup-kkmz1}
\bea
W &=& \sum_{i=1}^{3}\,[~\frac{1}{2}g_{i}^{u}
\,{\bf 10}_i{\bf 10}_i H_{i}+
g_{i}^{d}\,{\bf 10}_i \overline{\bf 5}_{i}\,
\overline{H}_{i}~] +
g_{23}^{u}\,{\bf 10}_2{\bf 10}_3 H_{4} \\
 & &+g_{23}^{d}\,{\bf 10}_2 \overline{\bf 5}_{3}\,
\overline{H}_{4}+
g_{32}^{d}\,{\bf 10}_3 \overline{\bf 5}_{2}\,
\overline{H}_{4}+
\sum_{a=1}^{4}g_{a}^{f}\,H_{a}\, 
{\bf 24}\,\overline{H}_{a}+
\frac{g^{\lambda}}{3}\,({\bf 24})^3~,\nonumber
\label{zoup-super}
\eea
where 
$H_{a}$ and $\overline{H}_{a}~~(a=1,\dots,4)$
stand for the Higgs quintets and anti-quintets.

The non-degenerate and isolated solutions to $\gamma^{(1)}_{i}=0$ for
the models $\{ {\bf A}~,~{\bf B} \}$ are: 
\bea (g_{1}^{u})^2
&=&\{\frac{8}{5},\frac{8}{5} \}g^2~, ~(g_{1}^{d})^2
=\{\frac{6}{5},\frac{6}{5}\}g^2~,~
(g_{2}^{u})^2=(g_{3}^{u})^2=\{\frac{8}{5},\frac{4}{5}\}g^2~,\label{zoup-SOL5}\\
(g_{2}^{d})^2 &=&(g_{3}^{d})^2=\{\frac{6}{5},\frac{3}{5}\}g^2~,~
(g_{23}^{u})^2 =\{0,\frac{4}{5}\}g^2~,~
(g_{23}^{d})^2=(g_{32}^{d})^2=\{0,\frac{3}{5}\}g^2~,
\nonumber\\
(g^{\lambda})^2 &=&\frac{15}{7}g^2~,~ (g_{2}^{f})^2
=(g_{3}^{f})^2=\{0,\frac{1}{2}\}g^2~,~ (g_{1}^{f})^2=0~,~
(g_{4}^{f})^2=\{1,0\}g^2~.\nonumber 
\eea 
According to the theorem of
ref.~\cite{zoup-lucchesi1} these models are finite to all orders.  After the
reduction of couplings the symmetry of $W$ is enhanced
\cite{zoup-finite1,zoup-kkmz1}.

The main difference of the models ${\bf A}$ and ${\bf B}$ is that
three pairs of Higgs quintets and anti-quintets couple to the ${\bf
  24}$ for ${\bf B}$ so that it is not necessary to mix them with
$H_{4}$ and $\overline{H}_{4}$ in order to achieve the triplet-doublet
splitting after the symmetry breaking of $SU(5)$.

In the dimensionful sector, the sum rule gives us the following
boundary conditions at the GUT scale \cite{zoup-kkmz1}:
\bea
m^{2}_{H_u}+
2  m^{2}_{{\bf 10}} &=&
m^{2}_{H_d}+ m^{2}_{\overline{{\bf 5}}}+
m^{2}_{{\bf 10}}=M^2~~\mbox{for}~~{\bf A} ~;\\
m^{2}_{H_u}+
2  m^{2}_{{\bf 10}} &=&M^2~,~
m^{2}_{H_d}-2m^{2}_{{\bf 10}}=-\frac{M^2}{3}~,~\nonumber\\
m^{2}_{\overline{{\bf 5}}}+
3m^{2}_{{\bf 10}}&=&\frac{4M^2}{3}~~~\mbox{for}~~{\bf B},
\eea
where we use as  free parameters 
$m_{\overline{{\bf 5}}}\equiv m_{\overline{{\bf 5}}_3}$ and 
$m_{{\bf 10}}\equiv m_{{\bf 10}_3}$
for the model ${\bf A}$, and 
$m_{{\bf 10}}\equiv m_{{\bf 10}_3}$  for ${\bf B}$, in addition to $M$.

\section{Predictions of Low Energy Parameters}
 
Since the gauge symmetry is spontaneously broken below $M_{\rm GUT}$,
the finiteness conditions do not restrict the renormalization properties
at low energies, and all it remains are boundary conditions on the
gauge and Yukawa couplings (\ref{zoup-SOL5}), the $h=-MC$ relation,
and the soft scalar-mass sum rule (\ref{zoup-sumr}) at $M_{\rm GUT}$,
as applied in the two models.  Thus we examine the evolution of
these parameters according to their RGEs up
to two-loops for dimensionless parameters and at one-loop for
dimensionful ones with the relevant boundary conditions.  Below
$M_{\rm GUT}$ their evolution is assumed to be governed by the MSSM.
We further assume a unique supersymmetry breaking scale $M_{s}$ (which
we define as the average of the stop masses) and
therefore below that scale the  effective theory is just the SM.

The predictions for the top quark mass $M_t$ are $\sim 183$ and $\sim 174$
GeV in models ${\bf A}$ and ${\bf B}$ respectively. Comparing these
predictions with the most recent experimental value $ M_t^{exp} =
(177.9\pm 4.4)$ GeV \cite{erler-PDG}, and recalling that the theoretical
values for $M_t$ may suffer from a correction of  $\sim 4 \%$
\cite{zoup-acta}, we see that they are consistent with the
experimental data.  In addition the value of $\tan \beta$ is found to be 
$\tan \beta \sim 54$ and $\sim 48$ for models ${\bf A}$ and ${\bf B}$ respectively.

\begin{figure}
           \centerline{\includegraphics[width=8cm,angle=-90]{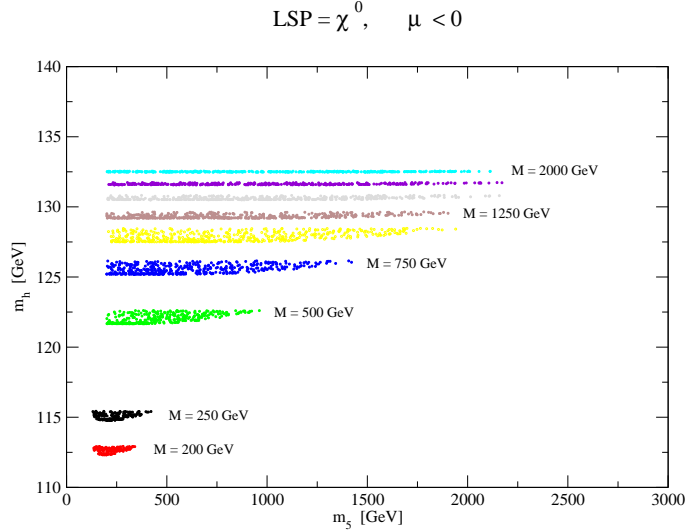}}
        \caption{$m_{h}$ as function of 
$m_{\bf 5}$ for different values of $M$ for model $\bf FUTA$, for $\mu<0 $.}
        %\label{zoup-fig:2}
        \end{figure}

In the SSB sector, besides the constraints imposed by finiteness there
are further restrictions imposed by phenomenology. In the case where
all the soft scalar masses are universal at the unfication scale,
there is no region of $M$ below $O(few~TeV)$ in which $m_{\tilde
  \tau} > m_{\chi^0}$ is satisfied (where $m_{\tilde \tau}$ is the lightest $\tilde \tau$
mass, and $m_{\chi^0}$ the lightest neutralino mass, which is the lightest
supersymmetric particle).  But once the universality condition is
relaxed this problem can be solved naturally (thanks to the sum rule).
More specifically, using the sum rule (\ref{zoup-sumr}) and imposing
the conditions a) successful radiative electroweak symmetry breaking,
b) $m_{\tilde\tau}^2>0$ and c) $m_{\tilde\tau}> m_{\chi^0}$, a comfortable
parameter space for both models (although model ${\bf B}$ requires
large $M\sim 1$ TeV) is found.

\begin{figure}
           \centerline{\includegraphics[width=8cm,angle=-90]{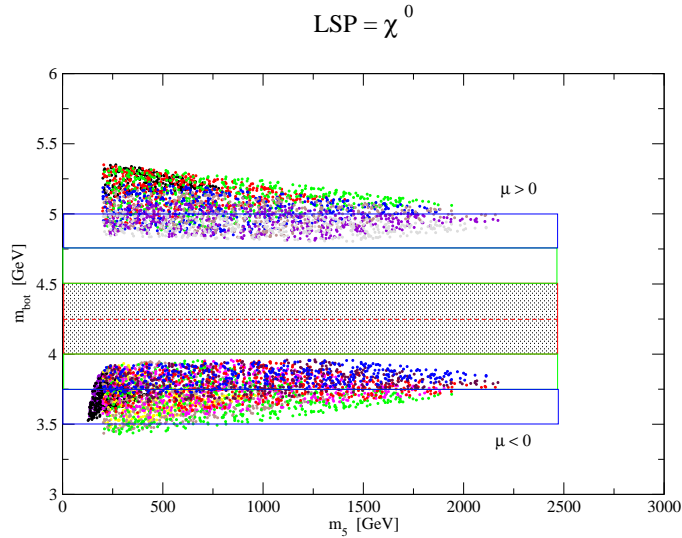}}
        \caption{$m_{bot}(m_{bot})$ as function of 
$m_{\bf 5}$ for different values of $M$ for model $\bf FUTA$, for
$\mu<0 $ and $\mu > 0$. The shaded region shows the experimentally
accepted value of $m_{bot}(m_{bot})$ according to ref.~\cite{PDG}. }
        %\label{zoup-fig:1}
        \end{figure}
\begin{figure}
           \centerline{\includegraphics[width=8cm]{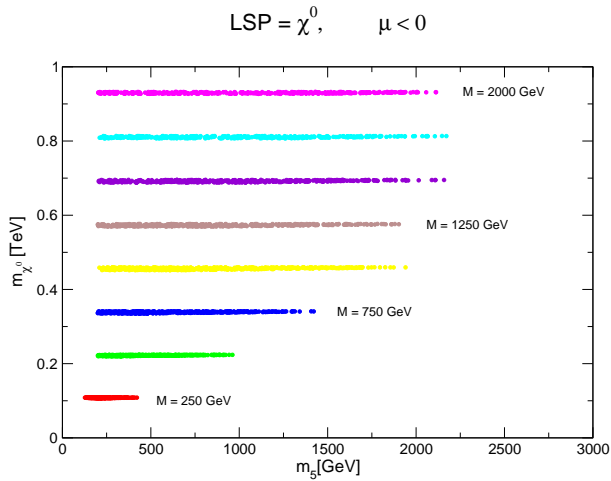}}
        \caption{$m_{\chi^0}$ as function of 
$m_{\bf 5}$ for different values of $M$ for model $\bf FUTA$, for $\mu<0 $.}
        %\label{zoup-fig:3}
        \end{figure}

As an additional constraint, we take into account the  $BR(b \to s \gamma)$
\cite{zoup-bsg}. %, whose experimental value is $1 \times 10^{-4} < BR(b \to s
%\gamma) < 4 \times 10^{-4}$. 
We do not take into account, though, constraints coming from the muon
anomalous magnetic moment (g-2) in this work, which excludes a small
region of the parameter space.  In the graphs we show the {\bf FUTA}
results concerning $m_h$, $m_{\chi^0}$, and $M_A$, for different values
of $M$, for the case where $\mu <0$ and the LSP is a neutralino $\chi^0$.
The results for $\mu>0$ are slightly different: the spectrum starts
around 500 GeV.  The main difference, though, is in the value of the
running bottom mass $m_{bot}(m_{bot})$, where we have included the
corrections coming from bottom squark-gluino loops and top
squark-chargino loops \cite{deltab}. In the $\mu <0$ case, $m_{bot} \sim
3.5-4.0~$ GeV is just below the experimental value $m_{bot}^{exp} \sim
4.0-4.5$ GeV \cite{PDG}, whereas in the $\mu >0$ case, $m_{bot} \sim 4.8
- 5.3$ GeV, i.~e.  above the experimental value.

\begin{figure}
           \centerline{\includegraphics[width=8cm]{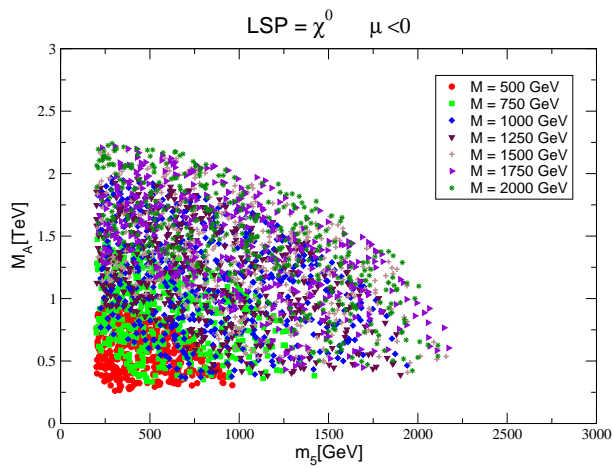}}
        \caption{$M_{A}$ as function of 
$m_{\bf 5}$ for different values of $M$ for model $\bf FUTA$, for $\mu<0 $.}
        %\label{zoup-fig:3}
\end{figure}

 \begin{table}
%\color{blue}\large
%\color{black}
\begin{center}
%\newcolumntype{M}{C@{\Cb\ (GeV)\quad}}
%\begin{tabular}{|M|C||M|C|}
\begin{tabular}{|c|c||c|c|}
 \hline
$M_{top}  $ & 183  & $m_{bot}  $ & 3.9  \\ \hline
$\tan\beta = $ & 54.4 & $\alpha_s $ & .118 \\ \hline\hline
$m_{\chi_1}$& 452 & $m_{\tilde{\tau}_2}$ & 916 \\ \hline
$m_{\chi_2}$& 843 & $m_{\tilde{\nu}_3}$ & 883 \\ \hline
$m_{\chi_3} $& 850 & $\mu$  & -1494 \\ \hline
$m_{\chi_4}$& 1500 & $B$  & 3543 \\ \hline
$m_{\chi^{\pm}_{1}} $ & 843 & $m_{A}$ & 555\\ \hline
$m_{\chi^{\pm}_{2}} $ & 1500  & $ m_{H^{\pm}}$ & 560 \\ \hline
$m_{\tilde{t}_1}$ & 1578 & $ m_{H}$ & 555 \\ \hline
$m_{\tilde{t}_2}$ & 1776 & $m_{h} $ & 127.5 \\ \hline
$m_{\tilde{b}_1}$ & 1580 & $M_1$ & 452 \\ \hline
$m_{\tilde{b}_2}$ & 1766 & $M_2 $ & 846  \\ \hline
$m_{\tilde{\tau}_1}$ & 654 & $ M_3 $ & 2210 
%\multicolumn{2}{l}{}
\\  \hline%\hhline{--~~}
\end{tabular}
\end{center}
\caption{A representative example of the bottom (running) and top
  (pole) masses, plus the supersymmetric spectrum for Model FUTA, with
  $m_5=$ 697 GeV, $m_{10}=$ 806 GeV,  $M_{susy}=$ 1681 GeV, $\mu < 0$.
  All masses in the Table are in GeV.} 
\end{table}

 \begin{table}
%\color{blue}\large
%\color{black}
\begin{center}
%\newcolumntype{M}{C@{\Cb\ (GeV)\quad}}
%\begin{tabular}{|M|C||M|C|}
\begin{tabular}{|c|c||c|c|}
 \hline
$M_{top}  $ & 173  & $m_{bot}  $ & 4.2  \\ \hline
$\tan\beta = $ & 48 & $\alpha_s $ & .116 \\ \hline\hline
$m_{\chi_1}$& 669 & $m_{\tilde{\tau}_2}$ & 970 \\ \hline
$m_{\chi_2}$& 912 & $m_{\tilde{\nu}_3}$ & 916 \\ \hline
$m_{\chi_3} $& 1289 & $\mu$  & -1900 \\ \hline
$m_{\chi_4}$& 1909 & $B$  & 4010 \\ \hline
$m_{\chi^{\pm}_{1}} $ & 1289 & $m_{A}$ & 1106 \\ \hline
$m_{\chi^{\pm}_{2}} $ & 909  & $ m_{H^{\pm}}$ & 1109 \\ \hline
$m_{\tilde{t}_1}$ & 2236 & $ m_{H}$ & 1106 \\ \hline
$m_{\tilde{t}_2}$ & 2519 & $m_{h} $ & 123.5 \\ \hline
$m_{\tilde{b}_1}$ & 2163 & $M_1$ & 700 \\ \hline
$m_{\tilde{b}_2}$ & 2501 & $M_2 $ & 1293  \\ \hline
$m_{\tilde{\tau}_1}$ & 766 & $ M_3 $ & 3256 
%\multicolumn{2}{l}{}
\\  \hline%\hhline{--~~}
\end{tabular}
\end{center}
\caption{A representative example of the bottom (running) and top
  (pole) masses, plus the supersymmetric spectrum for Model FUTB, with
  $m_{10}=$ 945 GeV,  $M_{susy}=$ 2278 GeV, $\mu < 0$.
  All masses in the Table are in GeV.} 
\end{table}

The Higgs mass prediction of the two models is, although the details
of each of the models differ, in the following range \beq m_h =\quad
\sim 112 - 132 ~~GeV, \eeq where the uncertainty comes from variations
of the gaugino mass $M$ and the soft scalar masses, and from finite
(i.e.~not logarithmically divergent) corrections in changing
renormalization scheme. The one-loop radiative corrections have been
included \cite{zoup-kaz-mh} for $m_h$, but not for the rest of the
spectrum. In making the analysis, the value of $M$ was varied from
$200-2000$ GeV for {\bf FUTA}. We have also included a small
variation, due to threshold corrections at the GUT scale, of $1-2 \%$
of the FUT boundary conditions.  This small variation does not give a
noticeable effect in the results at low energies. From Fig.~1 we
already see that the requirement $m_h > 114.4$ GeV \cite{X}
(neglecting the theoretical uncertainties) excludes the possibility of
$M=200$ GeV for FUTA, which is taken into account in the presentation
of the next graphs. In Tables 1 and 2 we present representative
examples of the values obtained for the s-spectra for each of the
models.

A more detailed numerical analysis, where the results of our program 
and of the known programs  FeynHiggs \cite{Heinemeyer:1998yj} and Suspect
\cite{Djouadi:2002ze} are combined, is currently in progress
\cite{zoup-abdelsven}. 

\section*{Acknowledgements}

We acknowledge very useful discussions with N.Benekos, G. Daskalakis,
R.Kinnunen and E. Richter-Was, as well as with J. Erler.  Supported by
the projects PAPIIT-IN116206 and Conacyt 42026-F, and partially
by the RTN contract HPRN-CT-2000-00148, the Greek-German Bilateral
Programme IKYDA-2001 and by the NTUA Programme for Fundamental
Research ``THALES''.

\end{document}